\documentclass[prd,reprint,superscriptaddress]{revtex4-1}

\usepackage{amsmath,amssymb}
\usepackage{verbatim}
\usepackage{graphicx}
\usepackage{hyperref}
\usepackage{color}
\usepackage{mathrsfs}
\usepackage{slashed}

\usepackage[utf8]{inputenc}

\newcommand {\cN}{{\cal N}}

\def\p{\pi}

\newcommand{\be}{\begin{equation}}
	\newcommand{\ee}{\end{equation}}
\newcommand{\bea}{\begin{eqnarray}}
	\newcommand{\eea}{\end{eqnarray}}

\newcommand{\ba}{\begin{array}}
	\newcommand{\ea}{\end{array}}

\def\double #1{#1{\hbox{\kern-2pt $#1$}}}

\newcommand{\bsubeq}{\begin{subequations}}
	\newcommand{\esubeq}{\end{subequations}}

\def\p{\psi}

\begin{document}
	
	\title{A Twisted Origin for Magnetic Carroll Supersymmetry}

		\author{Ilayda Bulunur}
	\email{bulunur@itu.edu.tr}
	\affiliation{Department of Physics,
		Istanbul Technical University,
		Maslak 34469 Istanbul,
		T\"urkiye}

		\author{Osman Ergec}
	\email{ergec24@itu.edu.tr}
	\affiliation{Department of Physics,
		Istanbul Technical University,
		Maslak 34469 Istanbul,
		T\"urkiye}

	\author{Oguzhan Kasikci}
	\email{kasikcio@itu.edu.tr}
	\affiliation{Department of Physics,
		Istanbul Technical University,
		Maslak 34469 Istanbul,
		T\"urkiye}
	
	\author{Mehmet Ozkan}
	\email{ozkanmehm@itu.edu.tr}
	\affiliation{Department of Physics,
		Istanbul Technical University,
		Maslak 34469 Istanbul,
		T\"urkiye}

		\author{Mustafa Salih Zog}
	\email{zog@itu.edu.tr}
	\affiliation{Department of Physics,
		Istanbul Technical University,
		Maslak 34469 Istanbul,
		T\"urkiye}

	\date{\today}
	
	%	\preprint{}

	\begin{abstract}
	Magnetic Carrollian theories provide a natural setting for field theories with nontrivial spatial structure in the Carroll limit and are therefore natural candidates for flat-space holographic duals. Embedding such boundary theories into a top-down framework requires a consistent supersymmetric completion and, in particular, an understanding of the relativistic origin of magnetic Carroll supersymmetry. We show that the relevant magnetic Carroll algebra does not arise from a naive contraction of the standard relativistic supersymmetry algebra, but instead descends from a twisted relativistic parent. As an explicit realization, we construct a three-dimensional $\cN=2$ magnetic Carroll algebra together with a supersymmetric vector-multiplet action. Unlike the electric case, the resulting structure contains one supercharge that squares to spatial momentum, a mixed anticommutator that yields the Hamiltonian, and a nilpotent second supercharge. We further show that its conformal extension coincides with the global part of a supersymmetric BMS$_4$ algebra. This provides a physical and relativistic origin for a super-BMS$_4$ structure recently identified by complementary algebraic methods, and strengthens the case for magnetic Carroll theories in flat-space holography and supersymmetric asymptotic symmetries.

	\end{abstract}
	
	%	\pacs{??? ... ???}
	
	\maketitle
	\allowdisplaybreaks

%\cite{Duval:2014uoa,  Hansen_2022, Bergshoeff:2017btm, Hartong:2015xda} supersymmetric theories   \cite {Ravera:2019ize, Ravera:2022buz, Zorba_2025}

	 \textit{Introduction} --
    More than half a century after its introduction \cite{Levy-Leblond:1965dsc, SenGupta:1966qer}, Carrollian physics has evolved from a largely formal construction into a framework of broad physical relevance \cite{Bagchi:2025vri}. While Carrollian symmetries have recently emerged in diverse physical settings, including, hydrodynamics \cite{ciambelli2018flat, Petkou:2022bmz, freidel2023carrollian, armas2024carrollian, ciambelli2018covariant}, condensed-matter systems with constrained mobility such as fractons \cite{ Bidussi:2021nmp, figueroa2023carroll, hartong2025fractons, Perez:2023uwt, Huang:2023zhp}, cosmological scenarios ranging from inflation to dark energy \cite{deBoer:2021jej}, and the near-horizon dynamics of black holes \cite{Penna:2018gfx, Donnay:2019jiz, ecker2023carroll, Redondo-Yuste:2022czg, gray2023carrollian, aggarwal2024carroll, Marsot:2022qkx, bivcak2023migrating}, their most far-reaching implications are expected to be in flat-space holography \cite{Bagchi:2016bcd, Donnay:2022aba, Bagchi:2022emh,donnay2023bridging, saha2023carrollian, nguyen2023carrollian, Ruzziconi:2026bix}. Holography in asymptotically flat spacetimes are expected to admit a boundary description governed by the symmetry structure of null infinity. In particular, the Bondi–Metzner–Sachs (BMS) group \cite{Bondi:1962px, Sachs:1962zza, McCarthy:1972} arises as the asymptotic symmetry group of four-dimensional flat spacetimes and may be understood as a conformal enhancement of Carrollian symmetries on a null boundary \cite{Duval:2014uva}. This viewpoint has motivated a Carrollian formulation of flat-space holography, according to which bulk quantum gravity is conjectured to be dual to a codimension-one Carrollian theory living at null infinity, much as conformal field theory provides the boundary description in AdS/CFT. From this viewpoint, it becomes important to understand which Carroll-invariant quantum field theories are viable candidates for boundary duals, and which structural features of the Carroll limit are compatible with a non-trivial holographic dictionary.
    
     Carroll symmetry arises as an ultra-relativistic contraction of relativistic kinematics, formally corresponding to the limit $c \to 0$.
     A salient consequence of the Carroll limit is the collapse of the light cone, which suppresses ordinary spatial propagation for timelike excitations. Upon explicitly reinstating the speed of light, time derivatives in relativistic actions are accompanied by factors of inverse power of $c$, rendering the strict $c \to 0$ limit singular. To obtain a finite Carrollian theory, one rescales the fields with powers of $c$ to remove these divergences. In standard two-derivative actions, this procedure leaves only the time-derivative terms at leading order, while spatial derivatives are suppressed or drop out entirely. The resulting dynamics is therefore ultralocal: spatial points decouple, and the theory reduces to an infinite collection of one-dimensional systems evolving independently in time. This phenomenon is illustrated sharply in the electric Carroll contraction of a free scalar, where the resulting theory can be viewed as an infinite number of decoupled quantum mechanical systems labeled by their spatial coordinates \cite{deBoer:2023fnj,Kasikci:2023zdn}. While ultralocality is a natural kinematic feature of the strict Carroll limit, it poses a conceptual challenge for holography. In continuum formulations, the absence of spatial derivatives generically enhances the symmetry: ultralocal Carrollian models are invariant under volume-preserving spatial diffeomorphisms (VPDs) \cite{Cotler:2024xhb}, i.e. coordinate transformations that preserve the spatial volume form. In this sense, the naive Carroll limit is “too symmetric” (and too ultralocal) to serve as a boundary dual with the correct set of symmetries.

     Several routes to nontrivial spatial structure are known. First, already at the level of Carroll contractions of Lorentz-invariant theories, there are inequivalent electric and magnetic limits \cite{Henneaux:2021yzg}, with the magnetic sector providing a natural arena for Carroll-invariant theories that evade strict ultralocality \cite{Cotler:2025npu}. Second, one may introduce higher-derivative interactions that preserve supertranslations while breaking the VPD symmetry. Upon quantization, however, such deformations appear to lead to spontaneous breaking of Carroll boosts. Alternatively, coupling matter to electric gauge fields can restore spatial dependence while preserving Carroll boosts \cite{Cotler:2024xhb}. This need for non-ultralocal physics becomes especially pressing in top-down approaches to Carrollian holography. In particular, a recent proposal has sought to define a Carrollian limit of the AdS/CFT correspondence, suggesting a duality between string theory in a Carrollian bulk geometry and the magnetic Carroll limit of $\cN=4$ Super Yang-Mills on the boundary \cite{Fontanella:2025tbs}. Such a proposal makes it essential to understand how fermions are realized in the magnetic Carroll sector and whether their couplings can be organized into fully consistent supersymmetric theories. This, in turn, raises a more basic question: what is the correct relativistic origin of magnetic Carroll supersymmetry?

     In this paper, we show that the supersymmetric magnetic Carroll sector is governed by an algebraic structure that does not arise from a naive contraction of the standard relativistic superalgebra. Rather, it originates from a twisted relativistic parent superalgebra of Hull type \cite{Hull:1998vg}, which we will refer to simply as a twisted ($\star$-) superalgebra, followed by a suitable redefinition of the supercharges and an ultra-relativistic contraction. This identifies the correct relativistic origin of magnetic Carroll supersymmetry and yields a systematic framework for constructing magnetic Carrollian field theories in which both bosons and fermions belong to the magnetic sector. As an explicit realization, we construct a three-dimensional $\cN = 2$ magnetic Carroll superalgebra and a supersymmetric vector-multiplet action. Three dimensions is especially natural for this purpose, since the conformal Carroll algebra in three-dimensions is isomorphic to the BMS$_4$ algebra, allowing us to directly infer the consequences of this magnetic structure for the asymptotic symmetries of four-dimensional flat spacetime. The resulting superalgebra differs sharply from the familiar electric case \cite{Bergshoeff:2015wma}: one supercharge squares to spatial momentum, the mixed anticommutator yields the Hamiltonian, and the second supercharge is nilpotent. The corresponding super-BMS$_4$ algebra has recently appeared in a complementary classification of electric and magnetic supersymmetric BMS$_4$ algebras \cite{Zheng:2025rfe}. Our result provides a concrete magnetic Carroll and twisted-relativistic origin for that structure. While our explicit construction is three dimensional, the underlying algebraic procedure is completely general.

\textit{Magnetic Carroll Supersymmetry --} A useful way to view magnetic Carroll theories is that, in Hamiltonian form, they retain the spatial part of the relativistic dynamics, while canonical momenta appear as Lagrange multipliers enforcing Carrollian constraints \cite{Henneaux:2021yzg}. For scalar and Maxwell fields, the magnetic Carroll limit preserves the spatial-gradient or magnetic sector, while the electric components are constrained. Relativistic fermions also admit electric and magnetic Carroll limits, and in the magnetic sector the equations of motion again impose a Carrollian time-independence constraint on-shell \cite{Bagchi:2022eui,Banerjee:2022ocj,Koutrolikos:2023evq,Bergshoeff:2023vfd}. Consequently, the magnetic sector retains nontrivial spatial structure for both bosons and fermions, while time evolution is constrained on-shell. In this sense, once the time-independence constraint is imposed, the magnetic Carroll sector closely parallels the structure of a Galilean theory, that is, the spatial dynamics remains nontrivial. This suggests that supersymmetric magnetic Carroll algebras should differ qualitatively from the familiar electric Carroll case, and should instead exhibit a structure closer to that of Galilean superalgebras, where supercharges square to spatial momentum.

To realize this expectation more concretely, here we focus on three-dimensional $\cN=2$ case. This example is particularly natural here, since its conformal extension admits an infinite-dimensional lift whose global part coincides with a supersymmetric BMS$_4$ algebra. Nevertheless, the mechanism to be discussed here is quite general. The minimal three-dimensional magnetic Carroll superalgebra is generated by time and space translations $(H,P_i)$, spatial rotations $J$, Carroll boosts $C_i$, and two Majorana supercharges $Q^\pm_\alpha$. The nonvanishing (anti)commutators are given by
\begin{align}
[J,Q^\pm_\alpha] &= -\frac12 (\gamma_0)_\alpha{}^\beta Q^\pm_\beta \,, & [J,C_i] &= \epsilon_{ij} C^j \,, \nonumber\\
\{Q^+_\alpha,Q^+_\beta\} &= -\frac12 (\gamma^i)_{\alpha\beta} P_i \,,  & [J,P_i] &= \epsilon_{ij} P^j  \,, \nonumber\\
\{Q^+_\alpha,Q^-_\beta\} &= \frac12 ({\gamma_0})_{\alpha\beta} H\,,& [C_i,P_j] &= - \epsilon_{ij} H \,, \nonumber\\
[C_i,Q^+_\alpha] &= -\frac12 (\gamma_i)_\alpha{}^\beta Q^-_\beta   \,, 
\label{MagneticCarrollSUSY}
\end{align}
This algebra was previously discussed in \cite{Zorba_2025} in the context of Carroll Galileons, and its two-dimensional analogue in \cite{Ravera:2022buz} in the context of Carroll Jackiw-Teitelboim supergravity.

The structure of \eqref{MagneticCarrollSUSY} makes this distinction from the electric case explicit. Rather than having supercharges that square only to the Hamiltonian, one supercharge squares to the spatial momentum, while the mixed anticommutator yields the Hamiltonian. This is precisely what one expects from the Galilean-type structure that emerges in the magnetic Carroll sector on-shell. It also explains why the minimal realization requires at least two supercharges: one must encode the nontrivial spatial sector, while the first-order time-derivative structure inherited from the relativistic parent still allows the Hamiltonian to appear in mixed anticommutators. This is reminiscent of extended non-relativistic superalgebras \cite{Bergman:1995zr,Bergman:1996bx,Bergshoeff:2014gja}, where multiple supercharges are likewise needed in order to accommodate both the Hamiltonian and spatial momentum within the supersymmetry algebra. A direct contraction of the standard Lorentzian relativistic $\cN=2$ superalgebra does not yield \eqref{MagneticCarrollSUSY} \footnote{One may formally recover the same algebra by passing to a Euclidean structure, assume real Grassmann variables, and imposing nonstandard reality conditions on the Grassmann bilinears \cite{Zorba_2025}}. Instead, the desired magnetic Carroll structure arises naturally from a twisted parent superalgebra 
\begin{align}
[J_A,Q^a_\alpha] &= -\frac12 (\gamma_A)_\alpha{}^\beta Q^a_\beta\,, & [J_A,X_B] &= \epsilon_{ABC} X^C  
 \,, \nonumber\\
\{Q^a_\alpha,Q^b_\beta\} &= -\frac12 \eta^{ab} (\gamma^A)_{\alpha\beta} P_A \,, 
\label{TwistedParent}
\end{align}
where $X_A = \{P_A, J_A\}$. Furthermore, $A=(0,i)$, $a,b=1,2$, and $\eta^{ab}=\mathrm{diag}(1,-1)$ encodes the twisted structure. 
Decomposing spacetime as $P_A = (P_0 \equiv H,\, P_i)\,,$ and $J_A = (J_0 \equiv J,\, J_i \equiv C_i)$, and redefining the supercharges according to \cite{LUKIERSKI2006389} 
\begin{equation}
Q^\pm_\alpha
=\frac{1}{\sqrt2}\Big(Q^1_\alpha \pm (\gamma_0)_\alpha{}^\beta Q^2_\beta\Big)\,,
\label{Qpmdef}
\end{equation}
we then perform the rescaling $H \rightarrow c^{-1} H\,, C_i \rightarrow c^{-1} C_i\,, Q^-_\alpha \rightarrow c^{-1} Q^-_\alpha\,,$ with $P_i$, $J$, and $Q^+_\alpha$ held fixed. In the $c\to0$ limit, \eqref{TwistedParent} reduces precisely to the magnetic Carroll superalgebra \eqref{MagneticCarrollSUSY}. The relative sign in $\eta^{ab}$ is exactly what allows the redefined supercharges to separate into a sector squaring to spatial translations and a mixed bracket yielding the Hamiltonian.

Having identified the relativistic origin of the magnetic Carroll superalgebra, we now turn to its minimal field-theoretic realization. But before doing so, a few technical notes are in order. A direct componentwise deformation of an ordinary $\cN=2$ Majorana multiplet from $SO(2)$ to $SO(1,1)$ is possible in principle, but it is not the most efficient route: once the internal metric is changed from $\delta_{ab}$ to $\eta_{ab}$, relative signs in the transformation rules and closure relations must be tracked carefully. A more practical formulation is obtained by rewriting the three-dimensional $\cN=2$ superalgebra and the supermultiplets using Dirac spinor with the following structure
\begin{equation}
   \psi=\psi_1+e\,\psi_2\,,\qquad e^2=+1\,, \qquad e^\star= - e\,,
\end{equation}
where $\p_{1,2}$ are Majorana spinors. Decomposing such a Dirac spinor back into Majorana components then yields a multiplet naturally adapted to the twisted superalgebra. As an example, consider a vector multiplet that consists of a real scalar field $\rho$, a $U(1)$ vector field $A_\mu$, a Dirac spinor $\lambda$ and an auxiliary scalar $D$. Writing the Dirac fermion as $\lambda=\lambda_1+e\,\lambda_2$, with $e^2=1$, and then decomposing back into Majorana components and passing to the $\pm$ basis in accordance with \eqref{Qpmdef}, one obtains the vector multiplet for the twisted superalgebra in $\pm$ basis where the anti-commutators take the following form
\begin{align}
\{Q_\pm, Q_\pm\} & = -\frac12 \gamma^i P_i \,,& \{Q_+, Q_-\} & =  \frac12 \gamma_0 H  \,.
\end{align}
The transformation rules for the vector multiplet in this basis is given by
\begin{align}
\delta A_0 &= \bar{\epsilon}_+\gamma_0 \lambda_- + \bar{\epsilon}_-\gamma_0 \lambda_+ \,,\nonumber\\
\delta A_i &= \bar{\epsilon}_+\gamma_i \lambda_+ + \bar{\epsilon}_-\gamma_i \lambda_- \,,\nonumber\\
\delta \rho &= 2\bar{\epsilon}_+\gamma_0 \lambda_+ - 2\bar{\epsilon}_-\gamma_0 \lambda_- \,,\nonumber\\
\delta \lambda_+ &= -\frac{1}{4}\gamma^{ij}F_{ij}\,\epsilon_+
-\frac{1}{2}\gamma^{0i}F_{0i}\,\epsilon_-
+\frac{1}{2}D\,\gamma_0\epsilon_-
 \nonumber\\
& -\frac{1}{4c}\dot{\rho}\,\epsilon_- -\frac{1}{4}\gamma_0{}^{i}\partial_i\rho\,\epsilon_+ \,,\nonumber\\
\delta \lambda_- &= -\frac{1}{4}\gamma^{ij}F_{ij}\,\epsilon_-
-\frac{1}{2}\gamma^{0i}F_{0i}\,\epsilon_+
-\frac{1}{2}D\,\gamma_0\epsilon_+
\nonumber\nonumber\\
&
+\frac{1}{4c}\dot{\rho}\,\epsilon_+ +\frac{1}{4}\gamma_0{}^{i}\partial_i\rho\,\epsilon_- \,,\nonumber\\
\delta D &= -\frac{1}{c}\bar{\epsilon}_+\dot{\lambda}_+
-\bar{\epsilon}_+\gamma_0{}^{i}\partial_i\lambda_-
+\bar{\epsilon}_-\gamma_0{}^{i}\partial_i\lambda_+ \nonumber\\
& +\frac{1}{c}\bar{\epsilon}_-\dot{\lambda}_- \,,
\label{TwistedVectorMultiplet}
\end{align}
where $F_{ij}=\partial_i A_j-\partial_j A_i$ and  $F_{0i}= c^{-1} \dot{A}_i-\partial_i A_0$. These rules realize the twisted relativistic parent multiplet in the $\pm$ basis. The transformations above leave invariant the relativistic twisted vector-multiplet action
\begin{align}
\mathcal{L}_{\rm V} &=
D^2+\frac{1}{4c^2}\dot{\rho}^{\,2}
-\frac14 \partial_i\rho\,\partial^i\rho
-\frac{2}{c}\bar\lambda_- {\gamma_0} \dot\lambda_+
 \nonumber\\
&-\frac{2}{c}\bar\lambda_+ {\gamma_0}  \dot\lambda_-
{+} 2\bar\lambda_+\gamma^i\partial_i\lambda_+
{+}2\bar\lambda_-\gamma^i\partial_i\lambda_- \nonumber\\
& +\frac12 F_{ij}F^{ij}+F_{0i}F^{0i}\,.
\label{TwistedRelAction}
\end{align}
While the relativistic twisted multiplet is off-shell, the magnetic Carroll limit is most naturally formulated in phase space. The canonical momenta then enter as additional variables whose supersymmetry transformations are fixed by the action through the supercharge. The resulting magnetic theory should therefore be viewed as a Hamiltonian realization of supersymmetry, rather than as an independent off-shell configuration-space multiplet. Having identified the twisted relativistic parent algebra, we now pass to its magnetic Carroll realization. 
To make this structure manifest, we first pass from the Lagrangian to its Hamiltonian variables by introducing the canonical momenta
\begin{equation}
E^{0i}\equiv \frac{\partial \mathcal L}{\partial \dot A_i}
= \frac{2}{c} F^{0i}\,,
\qquad
\pi \equiv \frac{\partial \mathcal L}{\partial \dot \rho}
= \frac1{2c^2}\dot\rho\,.
\end{equation}
We then rewrite the theory in first-order form and determine the supersymmetry transformations of the phase-space variables from their Poisson brackets with the conserved supercharges. With this preparation, we perform the magnetic Carroll scaling $\epsilon_- \to c\,\epsilon_-,\lambda_- \to c\,\lambda_-\,,
A_0 \to c^{-1} A_0\,, D \to c^2 D$ while keeping \(A_i\), \(\rho\), \(\lambda_+\), \(\pi\), and \(E_{0i}\) fixed, one finds that taking the $c\to 0$ limit yields
\begin{align}
\mathcal L_{\rm mag}
& = \pi \dot\rho
-\frac14 \partial_i\rho\,\partial^i\rho
+\frac12 F_{ij}F^{ij}
+E_{0i}F^{0i}
\nonumber\\
& -2 \bar\lambda_+ \gamma_0 \dot\lambda_- -2 \bar\lambda_- \gamma_0 \dot\lambda_+ +2\bar\lambda_+ \gamma^i \partial_i \lambda_+ \,,
\label{MagneticAction}
\end{align}
together with the supersymmetry transformations
\begin{align}
\delta \rho &= 2\bar\epsilon_+\gamma_0\lambda_+ \,,\nonumber\\
\delta A_0 &= 0 \,,\nonumber\\
\delta A_i &= \bar\epsilon_+\gamma_i\lambda_+ \,,\nonumber\\
\delta \lambda_+ &=
-\frac14 \gamma^{ij}F_{ij}\epsilon_+
-\frac14 \gamma_0{}^{i}\partial_i\rho\,\epsilon_+ \,,\nonumber\\
\delta \lambda_- &=
-\frac14 \gamma^{ij}F_{ij}\epsilon_-
-{\frac14 } \gamma^{0i}E_{0i}\epsilon_+
+\frac12 \pi\,\epsilon_+\nonumber\\
&
+\frac14 \gamma_0{}^{i}\partial_i\rho\,\epsilon_- \,,\nonumber\\
\delta \pi &=
\bar\epsilon_+\gamma^i\partial_i\lambda_-
-\bar\epsilon_-\gamma^i\partial_i\lambda_+ \,,\nonumber\\
\delta E_{0i} &= -{2}
\epsilon_{ij}\Big(\bar\epsilon_+\partial^j\lambda_-
+\bar\epsilon_-\partial^j\lambda_+\Big)\,.
\label{MagneticSUSY}
\end{align}
As expected, in this limit the conjugate momenta become Lagrange multipliers, imposing the constraints $\dot\phi  = 0 \,, \dot \lambda_+ = 0 \,,  F_{0i}  = 0 \,$. Furthermore, the auxiliary field $D$ decouples from the model. When the constraints are imposed, both the Lagrangian and the transformation rules resemble those of a Galilean model, which is possible because of the magnetic Carroll algebra. To our knowledge, \eqref{MagneticAction} provides the first realization of a magnetic Carroll supersymmetry whose origin can be traced to a relativistic parent superalgebra. The scaling chosen above leads to a fully magnetic supersymmetric Carroll model, in the sense that both the bosonic and fermionic sectors are of magnetic type. 

The Lagrangian \eqref{TwistedRelAction} admits another, hybrid limit where bosonic sector is electric, while the fermionic sector retains magnetic character. This is achieved by the alternative rescaling $ \epsilon_- \to c\,\epsilon_- \,, A_i \to c\,A_i \,, \lambda_+ \to c\,\lambda_+ \,, \rho \to c\,\rho $
while keeping the remaining fields and supersymmetry parameters fixed. In the $c\to0$ limit this yields
\begin{align}
\mathcal{L}_{\rm hybrid}
&=
D^2
+\frac14 \dot{\rho}^{\,2}
+F_{0i}F^{0i} -2\bar{\lambda}_-{\gamma_0}\dot{\lambda}_+ \nonumber\\
&
-2\bar{\lambda}_+{\gamma_0}\dot{\lambda}_-
{+}2\bar{\lambda}_-\gamma^i\partial_i\lambda_-\,.
\end{align}
The corresponding supersymmetry transformations follow from the same rescaling and the limit procedure.

\textit{Conformal Extension and Supersymmetric BMS$_4$}—
The twisted parent superalgebra admits a conformal extension which has the additional generators
$\{K_A,D,T,S^a_\alpha\}$,
where $K_A$ denote the special conformal generator, $D$ is the dilatation generator, $T$ is the $SO(1,1)$ R-symmetry generator, and $S^a_\alpha$ is the S-SUSY generator. Its explicit form, together with the twisted internal metric and the Carroll contraction, is given in the Supplemental Material. Here we instead focus on the basis in which the resulting conformal magnetic Carroll superalgebra coincides with the global part of a supersymmetric BMS$_4$ algebra. For the bosonic generators, we have the following definitions
\begin{align}
L_0 &= \tfrac12(D+iJ)\,, & L_{-1} &= \tfrac12(P_x+iP_y)\,, \nonumber\\
L_1 &= -\tfrac12(K_x-iK_y)\,,  & M_{-\frac12,-\frac12} &= -H\,, \nonumber\\
M_{-\frac12,\frac12} &= -C_y+iC_x\,, & M_{\frac12,\frac12} &= -K\,,  \nonumber\\
R_{0,0} &= -2T\,.
\label{BMSbosmap}
\end{align}
where $\bar{M}_{r,s} = M_{s,r}$ for the complex conjugation. For the fermionic sector, we have
\begin{align}
G_{-\frac12} &= \frac{e^{i\pi/4}}{\sqrt2}\left(Q^+_1-iQ^+_2\right)\,, \nonumber\\
G_{+\frac12} &= \frac{e^{3i\pi/4}}{\sqrt2}\left(S^-_1+iS^-_2\right)\,, \nonumber\\
N_{0,-\frac12} &= \sqrt2\,e^{i\pi/4}\left(Q^-_1+iQ^-_2\right)\,, \nonumber\\
N_{0,+\frac12} &= -\sqrt2\,e^{3i\pi/4}\left(S^+_1-iS^+_2\right)\,, \nonumber\\
\label{BMSfermmap}
\end{align}
In terms of these generators, and working in a really real representation of the gamma matrices with
$\gamma_0=i\sigma_2$, $\gamma_1=\sigma_1$, and $\gamma_2=\sigma_3$,
the conformal magnetic Carroll superalgebra takes the form of the global part of the following supersymmetric BMS$_4$ algebra
\begin{widetext}
\begin{equation}
\begin{alignedat}{3}
[L_n,L_m] &= (n-m)L_{n+m}, \qquad&
[L_m,M_{r,s}] &= \Bigl(\frac{m}{2}-r\Bigr) M_{m+r,s}, \qquad&
\{G_r,G_s\} &= L_{r+s}, \\[2pt]
\{G_r,N_{m,s}\} &= M_{r+m,s}, &
[L_n,G_r] &= \Bigl(\frac{n}{2}-r\Bigr) G_{n+r}, &
[M_{r,s},G_p] &= \frac12 (r-p)N_{r+p,s}, \\[2pt]
[L_n,N_{k,s}] &= -k\,N_{n+k,s}, &
\{G_r,\bar N_{k,s}\} &= \frac12 (r-s)R_{k,r+s}, &
[\bar L_n,N_{k,r}] &= \Bigl(\frac{n}{2}-r\Bigr) N_{k,n+r}, \\[2pt]
[R_{k,n},G_r] &= -\bar N_{k,r+n}, &
[L_k,R_{m,n}] &= -n\,R_{m,k+n}\,, &&
\end{alignedat}
\label{eq:superbms}
\end{equation}
\end{widetext}
together with their complex conjugates. The BMS$_4$ superalgebra is generated by
$L_n$, $G_p$, $M_{r,s}$, $N_{k,p}$, and $R_{k,n}$, with
$m,n,k,l\in\mathbb Z$ and $r,s,p\in\mathbb Z+\frac12$.
Barred generators denote the complex conjugates of their unbarred counterparts, with $\bar R_{k,n}=R_{n,k}$. Supersymmetric BMS$_4$ algebras have been studied previously in \cite{Awada:1985by,Fotopoulos:2020bqj,Henneaux:2020ekh,Fuentealba:2021xhn,Bagchi:2022owq,Zheng:2025cuw,Zheng:2025rfe}. In particular, the algebra obtained here coincides with the Type I-I magnetic super-BMS$_4$ algebra identified in the recent classification of \cite{Zheng:2025rfe}. Our result provides a twisted relativistic Lorentzian parent for this algebra, and hence a physical origin for this class of magnetic super-BMS$_4$ symmetry.

 \textit{Outlook}.—
In this paper, we presented the first example of a supersymmetric magnetic Carroll action with a clear relativistic origin. At the algebraic level, we showed that the magnetic superalgebra admits a conformal extension that is the global part of a supersymmetric BMS$_4$ algebra. This identifies a concrete magnetic Carroll and twisted-relativistic origin for a class of magnetic super-BMS$_4$ symmetries, and suggests that such algebras are not merely formal possibilities but can arise naturally from relativistic parent structures.

A natural next step is to determine asymptotic boundary conditions in four-dimensional supergravity that realize this class of magnetic super-BMS$_4$ symmetry as an asymptotic symmetry algebra. This would clarify whether the magnetic supersymmetric structure identified here has a direct bulk interpretation, and would place it on the same footing as more familiar asymptotic symmetry constructions \cite{Awada:1985by}. More generally, it would be interesting to understand whether twisted relativistic parent algebras provide a systematic route to magnetic asymptotic symmetry algebras in flat space, rather than merely isolated examples.

On the field-theory side, it would be interesting to extend the present abelian model to interacting, and to non-abelian and higher-dimensional models, in particular to $\cN=4$ SYM. Such extensions could help clarify whether the twisted relativistic construction developed here persists in genuinely interacting settings, and whether it can provide a natural framework for supersymmetric magnetic Carroll matter and gauge multiplets beyond the minimal example studied in this work. It would also be interesting to understand to what extent the magnetic construction can be implemented off-shell for extended supersymmetry when both the bosonic and the fermionic sectors are magnetic.

More broadly, our results suggest that twisted relativistic parent algebras may provide a concrete route into supersymmetric Carrollian holography, where magnetic super-BMS$_4$ algebras, Carroll geometry in flat-space or de Sitter settings, and Carrollian limits of gauge/string dualities have all begun to appear in recent work \cite{Blair:2025nno, Gomis:2023eav, Fontanella:2025tbs}. In particular, our results suggest that twisted relativistic supersymmetry may provide a natural framework for magnetic Carroll theories and their flat-space holographic interpretation.

	\textit{Acknowledgements.} -- We are grateful to Utku Zorba for feedback and comments on the manuscript. M.O. is supported in part by TUBITAK grant 125F471 and Istanbul Technical University Research Fund under grant number TGA-2025-46976. M.O. also acknowledges support from the TÜBİTAK Incentive Award. M.S.Z. and I.B. acknowledge support from the Istanbul Technical University Research Fund under grant numbers TDK-2025-47808 and TDK-2026-48231, respectively.

	\bibliographystyle{utphys}
	\bibliography{ref}
 \clearpage

\appendix{}

\begin{center}\textbf{SUPPLEMENTAL MATERIAL\\
(APPENDICES)}\end{center}

In this Supplemental Material we present the explicit twisted conformal parent superalgebra, its spacetime decomposition, and the Carroll contraction leading to the conformal magnetic Carroll superalgebra used in the main text. 

\section{Twisted conformal parent algebra and Conformal Magnetic Carroll Contraction}
The twisted three-dimensional $\cN=2$ conformal superalgebra is generated by
$\{J_A,P_A,K_A,D,T,Q^a_\alpha,S^a_\alpha\}$, with
$A=0,1,2$, $a=1,2$, and twisted internal metric
$\eta^{ab}={\textrm{diag}}(1,-1)$. Its nonvanishing bosonic commutators are
\begin{align}
[J_A,P_B] &= \epsilon_{ABC}P^C, &
[J_A,K_B] &= \epsilon_{ABC}K^C, \nonumber\\
[D,K_A] &= -K_A, &
[P_A,K_B] &= 2\eta_{AB}D+2\epsilon_{ABC}J^C,\nonumber\\
[J_A,J_B] &= \epsilon_{ABC}J^C, &
[D,P_A] &= P_A\,.
\end{align}
while the non-vanishing bose-fermi commutators are given by
\begin{align}
[D,Q^a_\alpha] &= \tfrac12 Q^a_\alpha, &
[D,S^a_\alpha] &= -\tfrac12 S^a_\alpha, \nonumber\\
[J_A,Q^a_\alpha] &= -\tfrac12(\gamma_A)_\alpha{}^\beta Q^a_\beta, &
[J_A,S^a_\alpha] &= -\tfrac12(\gamma_A)_\alpha{}^\beta S^a_\beta, \nonumber\\
[P_A,S^a_\alpha] &= (\gamma_A)_\alpha{}^\beta Q^a_\beta, &
[K_A,Q^a_\alpha] &= (\gamma_A)_\alpha{}^\beta S^a_\beta, \nonumber\\
[T,Q^a_\alpha] &= \epsilon^{ab}\eta_{bc}Q^c_\alpha, &
[T,S^a_\alpha] &= \epsilon^{ab}\eta_{bc}S^c_\alpha,
\end{align}
and the anticommutators are 
\begin{align}
\{Q^a_\alpha,Q^b_\beta\}
&= -\tfrac12 \eta^{ab}(\gamma_A)_{\alpha\beta}P^A, \nonumber\\
\{Q^a_\alpha,S^b_\beta\}
&= -\tfrac12 \eta^{ab} C_{\alpha\beta}D
+\tfrac12 \eta^{ab}(\gamma_A)_{\alpha\beta}J^A
+\tfrac12 \epsilon^{ab} C_{\alpha\beta}T \,, \nonumber\\ 
\{S^a_\alpha,S^b_\beta\}
&= -\tfrac12 \eta^{ab}(\gamma_A)_{\alpha\beta}K^A \,.
\end{align}

To take the Carroll limit to obtain a magnetic superconformal Carroll algebra, we first we decompose $P_A, J_A,K_A$ as $P_A=(H,P_i)\,, J_A=(J,C_i)\,, K_A=(K,K_i)$ and define 
\begin{align}
Q^\pm_\alpha
&=\frac1{\sqrt2}\Bigl(Q^1_\alpha
\pm (\gamma_0)_\alpha{}^\beta Q^2_\beta\Bigr), \nonumber\\
S^\pm_\alpha
&=\frac1{\sqrt2}\Bigl(S^1_\alpha
\pm (\gamma_0)_\alpha{}^\beta S^2_\beta\Bigr).
\end{align}
Next, we introduce the $c$-scalings of the generators. In the superconformal case, $
H,K,C_i,Q^-_\alpha,S^+_\alpha,T$ are scaled with the inverse power of c
while keeping $P_i$, $J$, $K_i$, $D$, $Q^+_\alpha$, and $S^-_\alpha$
fixed. Finally, we take $c\to0$ limit, which yields the conformal magnetic
Carroll superalgebra. Its nonvanishing bosonic commutators are given by
\begin{align}
[J,P_i] &= \epsilon_{ij}P^j, &
[J,C_i] &= \epsilon_{ij}C^j, \nonumber\\
[J,K_i] &= \epsilon_{ij}K^j, &
[C_i,P_j] &= -\epsilon_{ij}H, \nonumber\\
[C_i,K_j] &= -\epsilon_{ij}K, &
[D,H] &= H, \nonumber\\
[D,P_i] &= P_i, &
[D,K] &= -K, \nonumber\\
[D,K_i] &= -K_i, &
[H,K_i] &= 2\epsilon_{ij}C^j, \nonumber\\
[P_i,K] &= -2\epsilon_{ij}C^j, &
[P_i,K_j] &= 2\delta_{ij}D-2\epsilon_{ij}J .
\end{align}
while the non-vanishing bose-fermi commutators are
\begin{align}
[D,Q^\pm_\alpha] &= \tfrac12 Q^\pm_\alpha, &
[D,S^\pm_\alpha] &= -\tfrac12 S^\pm_\alpha, \nonumber\\
[J,Q^\pm_\alpha] &= -\tfrac12(\gamma_0)_\alpha{}^\beta Q^\pm_\beta, &
[J,S^\pm_\alpha] &= -\tfrac12(\gamma_0)_\alpha{}^\beta S^\pm_\beta, \nonumber\\
[C_i,Q^+_\alpha] &= -\tfrac12(\gamma_i)_\alpha{}^\beta Q^-_\beta, &
[C_i,S^-_\alpha] &= -\tfrac12(\gamma_i)_\alpha{}^\beta S^+_\beta, \nonumber\\
[K,Q^+_\alpha] &= (\gamma_0)_\alpha{}^\beta S^+_\beta, &
[K_i,Q^+_\alpha] &= (\gamma_i)_\alpha{}^\beta S^-_\beta, \nonumber\\
[H,S^-_\alpha] &= (\gamma_0)_\alpha{}^\beta Q^-_\beta, &
[P_i,S^+_\alpha] &= (\gamma_i)_\alpha{}^\beta Q^-_\beta, \nonumber\\
[T,Q^+_\alpha] &= -(\gamma_0)_\alpha{}^\beta Q^-_\beta, &
[T,S^-_\alpha] &= (\gamma_0)_\alpha{}^\beta S^+_\beta, \nonumber\\
[K_i,Q^-_\alpha] &= (\gamma_i)_\alpha{}^\beta S^+_\beta, &
[P_i,S^-_\alpha] &= (\gamma_i)_\alpha{}^\beta Q^+_\beta ,
\end{align}
Finally, the non-vanishing anticommutator structure is given by
\begin{align}
\{Q^+_\alpha,Q^+_\beta\}
&= -\tfrac12(\gamma^i)_{\alpha\beta}P_i, \nonumber\\
\{Q^+_\alpha,Q^-_\beta\}
&= -\tfrac12(\gamma^0)_{\alpha\beta}H, \nonumber\\
\{S^-_\alpha,S^-_\beta\}
&= -\tfrac12(\gamma^i)_{\alpha\beta}K_i, \nonumber\\
\{S^+_\alpha,S^-_\beta\}
&= -\tfrac12(\gamma^0)_{\alpha\beta}K, \nonumber\\
\{Q^+_\alpha,S^-_\beta\}
&= -\tfrac12 C_{\alpha\beta}D
-\tfrac12(\gamma_0)_{\alpha\beta}J, \nonumber\\
\{Q^-_\alpha,S^-_\beta\}
&= \tfrac12 C_{\alpha\beta}T
+\tfrac12(\gamma^i)_{\alpha\beta}C_i, \nonumber\\
\{Q^+_\alpha,S^+_\beta\}
&= -\tfrac12(\gamma_0)_{\alpha\beta}T
+\tfrac12(\gamma^i)_{\alpha\beta}C_i .
\end{align}
This is the three-dimensional $\cN = 2$ superconformal magnetic Carroll superalgebra.

\end{document}